\documentclass[twocolumn,aps,amssymb,showpacs,superscriptaddress,secnumarabic]{revtex4}
\usepackage{graphicx}
\usepackage{amsmath}
\usepackage{mathtools}
\usepackage{chngcntr}
\usepackage{bm}
\graphicspath{%
    {converted_graphics/}
    {/}
}
\begin{document}
\date{\today}
\title {Theory of the Transmission of Infection in the Spread of Epidemics:\\ Interacting Random Walkers with and without Confinement}

\author{V. M. Kenkre} 
\email{kenkre@unm.edu}
\author{S. Sugaya} 
\email{satomi@unm.edu}
\affiliation{Consortium of the Americas for Interdisciplinary Science and the Department of Physics and Astronomy, University of New Mexico, Albuquerque, New Mexico 87131}

\begin{abstract}
A theory of the spread of epidemics is formulated on the basis of pairwise interactions in a dilute system of random walkers (infected and susceptible animals) moving in $n$ dimensions. The motion of an animal pair is taken to obey a Smoluchowski equation in 2n-dimensional space that combines diffusion  with confinement of each animal to its particular home range. An additional (reaction) term that comes into play when the animals are in close proximity describes the process of infection. Analytic solutions are obtained, confirmed by numerical procedures, and shown to predict a surprising effect of confinement. The effect is that infection spread has a non-monotonic dependence on the diffusion constant and/or the extent of the attachment of the animals to the home ranges. Optimum values of these parameters exist for any given distance between the attractive centers. Any change from those values, involving faster/slower diffusion or shallower/steeper confinement, hinders the transmission of infection. A physical explanation is provided by the theory. Reduction to the simpler case of no home ranges is demonstrated. Effective infection rates are calculated  and it is shown how to use them in  complex systems consisting of dense  populations.
\end{abstract}
. 

\maketitle

The purpose of the following is to construct an analytic theory of the transmission of infection in epidemics spread on the basis of a simple but exactly soluble model of interacting random walkers representing animals moving about on the terrain and infecting one another on encounter. Seminal contributions by Anderson and May and others~\cite{may, OkuboLevin, Hethcote, BAUER}, involving concepts such as mass action, SIR, and the basic reproductive rate $R_0,$ launched this field of research which derives its importance from human relevance as well as intellectual challenge. Spatial considerations were introduced into the investigations independently by various authors~\cite{OkuboLevin,DLM:2000,AK:2002,CantrellCosner, KENKRE215:219,McKane,KENKRE203:233,AGUIRRE,McIinnisThesis} giving the studies a kinetic equation flavor. Missing from some of these studies were confinement features that arise in animal motion from home ranges and yet are clear and compelling in the light of field observations~\cite{GAKSMY:2005,jtb,McIinnisThesis}. These and other issues have made it essential to undertake a fundamental study of the transmission of infection in terms of interacting random walks specially under confinement.

\emph{Model and Method of Analysis--} Our model starts with just two animals, one initially infected and the other initially uninfected (susceptible), respectively denoted by $1$ and $2$, performing random walks around respective attractive centers at $\bm{R}_{1}$ and $\bm{R}_{2}$, with a diffusion constant $D$, there being the possibility of the uninfected individual getting infected at a rate proportional to $\mathcal{C}$ when the two occupy the same position. The central quantity that serves as the focus of our calculation is the joint probability density $P(\bm{r}_1,\bm{r}_2,t)$ that the infected animal is at $\bm{r}_1$ and the susceptible animal is at $\bm{r}_2$. Given this definition,  $P(\bm{r}_1,\bm{r}_2,t)$ vanishes when the susceptible animals gets infected and the infection problem becomes formally similar to a Frenkel exciton annihilation problem analyzed a number of years ago~\cite{vmk80}. The present problem is considerably more complex, however, as a consequence of the tethering of the individuals to separate  centers. Guided by the procedures set out in reference~\cite{vmk80}, we consider a capture problem in a space of twice the number of dimensions as the space in which each walker moves, introduce attractive quadratic potentials of steepness $\gamma$ around the centers at $\bm{R}_{1}$ and $\bm{R}_{2}$, and  write, applicable to $s$-dimensions in general,
\begin{align}
\frac{\partial P}{\partial t} = & \nabla_{1}\cdot\left[\gamma \left(\bm{r}_{1} - \bm{R}_{1}\right)P\right]  + \nabla_{2}\cdot\left[\gamma \left(\bm{r}_{2} - \bm{R}_{2}\right)P\right]  \nonumber \\
& + D\left(\nabla_1^2+\nabla_2^2\right)P -\delta(\bm{r}_1-\bm{r}_2) \mathcal{C}P.
\label{start}
\end{align}
In terms of the propagator (Green function) for the homogeneous problem, $\Pi(\bm{r}_1,\bm{r}_1^0,\bm{r}_2,\bm{r}_2^0,t)$, the solution in the absence of the infection rate for any initial placement of the two animals given by $P(\bm{r}_1,\bm{r}_2,0)$ would be
\begin{equation}
 \eta(\bm{r}_1,\bm{r}_2,t)=\int_{-\infty}^{\infty}\int_{-\infty}^{\infty}d^sr_1^0 d^sr_2^0 \,\,\Pi(\bm{r}_1,\bm{r}_1^0,\bm{r}_2,\bm{r}_2^0,t)P(\bm{r}_1^0,\bm{r}_2^0,0).
\end{equation}
When infection is present, we write, as a consequence of the linearity of the equations,
\begin{align}
& P(\bm{r}_1,\bm{r}_2,t)  = \eta(\bm{r}_1,\bm{r}_2,t)\nonumber \\
& -\mathcal{C}\int_0^t dt'\int_{-\infty}^{\infty}d^sr'_1\,\,\Pi(\bm{r}_1,\bm{r}'_1,\bm{r}_2,\bm{r}'_1,t-t')P(\bm{r}'_1,\bm{r}'_1,t').
\label{withinfection}
\end{align}
Defect technique procedures \cite{montroll,Katja3,KENKRE:1983,SZABO:1984:ID171,KSinJPC,REDNER:2001:ID188,SSK:2013,vmknu} along the lines originated in \cite{vmk80}, proceed by Laplace transforming Eq. (\ref{withinfection}), setting $\bm{r}_1=\bm{r}_2,$ and integrating over $\bm{r}_1$ in the appropriate space of $s$ dimensions. An important result is 
\begin{align}
& \int_{-\infty}^{\infty}d^sr_1\,\,\tilde{P}(\bm{r}_1,\bm{r}_1,\epsilon)= \int_{-\infty}^{\infty}d^sr_1 \,\,\tilde{\eta}(\bm{r}_1,\bm{r}_1,\epsilon) -\mathcal{C}\,\times\nonumber \\
&\int_{-\infty}^{\infty}d^sr'_1 \int_{-\infty}^{\infty}d^sr_1\,\,\tilde{\Pi}(\bm{r}_1,\bm{r}'_1,\bm{r}_1,\bm{r}'_1,\epsilon)\tilde{P}(\bm{r}'_1,\bm{r}'_1,\epsilon),
\label{prenu}
\end{align}
where $\epsilon$ is the Laplace variable and tildes denote Laplace transforms. Motivated by the so-called nu-function analysis introduced in capture problems \cite{vmknu} (for a recent review and application see \cite{KSinJPC}), and assisted by the observation that the integral of $\tilde{\Pi}(\bm{r}_1,\bm{r}'_1,\bm{r}_1,\bm{r}'_1,\epsilon)$ over the entire domain of $\bm{r}_1$  (i.e., all space) appearing in Eq. (\ref{prenu}) is independent of $\bm{r}'_1$, we introduce the  symbol $ \tilde{\nu}(\epsilon)$ to denote that integral,
\begin{equation}
 \tilde{\nu}(\epsilon) = \int_{-\infty}^{\infty}d^sr_1\,\,\tilde{\Pi}(\bm{r}_1,\bm{r}'_1,\bm{r}_1,\bm{r}'_1,\epsilon),
 \label{nutilde}
\end{equation}
and succeed in obtaining, in the Laplace domain, an \emph{explicit} solution for the joint probability (density) that the two animals occupy the same position,
\begin{equation}
  \int_{-\infty}^{\infty}d^sr'_1\,\,\tilde{P}(\bm{r}'_1,\bm{r}'_1,\epsilon)=\frac{\tilde{\mu}(\epsilon)}{1+\mathcal{C}\tilde{\nu}(\epsilon)}.
 \label{usingmunu}
\end{equation}

The expression in Eq.~(\ref{usingmunu}) contains two quantities that are key to the analysis. The first of these, $\nu(t)$, whose Laplace transform is defined in Eq.~(\ref{nutilde}), is the probability (density) that the locations of the two animals coincide (whatever that location) if at a time $t$ earlier their locations also coincided. The second key quantity, $\mu(t)$, whose Laplace transform is
\begin{align}
 & \tilde{\mu}(\epsilon) = \int_{-\infty}^{\infty}d^sr'_1\,\,\tilde{\eta}(\bm{r}'_1,\bm{r}'_1,\epsilon)=\int_{-\infty}^{\infty}d^sr'_1\,\,\times \nonumber \\
 & \int_{-\infty}^{\infty}\int_{-\infty}^{\infty}d^sr_1^0 d^sr_2^0\,\,\tilde{\Pi}(\bm{r}'_1,\bm{r}_1^0,\bm{r}'_1,\bm{r}_2^0,\epsilon)P(\bm{r}_1^0,\bm{r}_2^0,0),
 \label{mutilde}
\end{align}
is the probability (density) that the two animals occupy the same location at the present time (whatever that location) if at a time $t$ earlier they occupied locations as per the \emph{given initial condition} of the problem. Both refer to the problem without infection ($\mathcal{C}=0$). They are integrals (over the $s$-dimensional space) of the two-particle joint probability density and have the dimensions of reciprocal length raised to $s$.
The rest of the calculation is straightforward. Knowledge of the propagators of the system generally in the presence of constraining potentials gives $\nu$ and, in combination with the given initial conditions, yields $\mu$. The two together with Eq.~(\ref{usingmunu}) provide all that is necessary to obtain the infection probability and the nuances of its behavior.

 \emph{Infection Curve and its Nonmonotonic Dependence}--When a definite infection event occurs,  the joint probability density $P(\bm{r}_1,\bm{r}_2,t)$ drops to zero. The infection probability is, therefore, 
\begin{equation}
 \mathcal{I}(t) = 1 - \int_{-\infty}^{\infty}\int_{-\infty}^{\infty}d^sr_1d^sr_{2}\,\,P(\bm{r}_1,\bm{r}_2,t),
\label{it}
\end{equation}
and, from Eq. (\ref{withinfection}), is obtained in the Laplace domain as
\begin{equation}
  \tilde{\mathcal{I}}(\epsilon) = \frac{1}{\epsilon}\left[\frac{\tilde{\mu}(\epsilon)}{(1/\mathcal{C})+\tilde{\nu}(\epsilon)}\right].
  \label{iepsilon}
 \end{equation}
Further insight requires the evaluation of the key quantities $\mu$ and $\nu$, which follows from the form of the propagators appropriate to Eq.~(\ref{start}). These are well-known to be Gaussian, to be multiplicative in Cartesian coordinates as one proceeds to higher dimensions, and to involve the saturating time $\mathcal{T}(t) = (1/2\gamma)(1-e^{-2\gamma t})$ 
that emerges from standard Ornstein-Uhlenbeck arguments~\cite{RISKEN:1989:ID199}. The $2s$-dimensional propagator and the resulting $\nu$ and $\mu$ functions, the latter for arbitrary initial placement, are 

\begin{align}
&\Pi(\bm{r}_1,\bm{r}_1^0,\bm{r}_2,\bm{r}_2^0,t) = \left(\frac{1}{4\pi D \mathcal{T}(t)}\right)^s\,\,\times\nonumber\\
&\prod_{\beta=1}^{s}e^{-\frac{\left(x_1^{\beta}-h_1^{\beta}-(x_{1}^{0\beta}-h_1^{\beta})e^{-\gamma t}\right)^2+\left(x_2^{\beta}-h_2^{\beta}-(x_{2}^{0\beta}-h_2^{\beta})e^{-\gamma t}\right)^2}{4D\mathcal{T}(t)}},\nonumber \\
&\nu(t) =\left(\frac{1}{\sqrt{8\pi D\mathcal{T}(t)}}\right)^s\prod_{\beta=1}^{s}e^{-\frac{\left(1-e^{-\gamma t}\right)^2\left(h_{1}^{\beta}-h_{2}^{\beta}\right)^2}{8D\mathcal{T}(t)}},\nonumber\\
 &\mu(t) = \left(\frac{1}{\sqrt{8\pi D\mathcal{T}(t)}}\right)^s\prod_{\beta=1}^{s}e^{-\frac{\left(h_{1}^{\beta}-h_{2}^{\beta}+\left(\left(x_1^{0\beta}-h_1^{\beta}\right)-\left(x_2^{0\beta}-h_2^{\beta}\right)\right)e^{-\gamma t}\right)^2}{8D\mathcal{T}(t)}}, 
\label{crux}
\end{align}
where the label $\beta$ runs from $1$ to $s$, and the initial position and home range center of the susceptible animal have the respective $x-$components $x_2^{0\beta}$ and $h_2^{\beta}$. The rest of the notation is obvious.

For the motion of two 1-dimensional walkers ($s=1$), we do not need the index $\beta$ and, if we make the natural assumption that the animals are located initially at their own respective centers, the quantities $\nu(t)$, $\mu(t)$, which are closely related to Smoluchowski propagators connecting the two home range centers,  are given  by
\begin{equation}
\nu(t) =\frac{e^{-\frac{H^2}{8D\mathcal{T}(t)}\left(1-e^{-\gamma t}\right)^2}}{\sqrt{8\pi D\mathcal{T}(t)}};\quad
 \mu(t) = \frac{e^{-\frac{H^2}{8D\mathcal{T}(t)}}}{\sqrt{8\pi D\mathcal{T}(t)}}.
 \label{numut}
\end{equation}
They equal each other for large times but begin quite differently at the initial time: $\mu(0)$ vanishes while $\nu(0)$ is infinite. Here $H=h_{1}-h_{2}$ is the distance between the two home range centers.

\begin{figure}[h] 
  \centering
   \includegraphics[width=1\columnwidth]{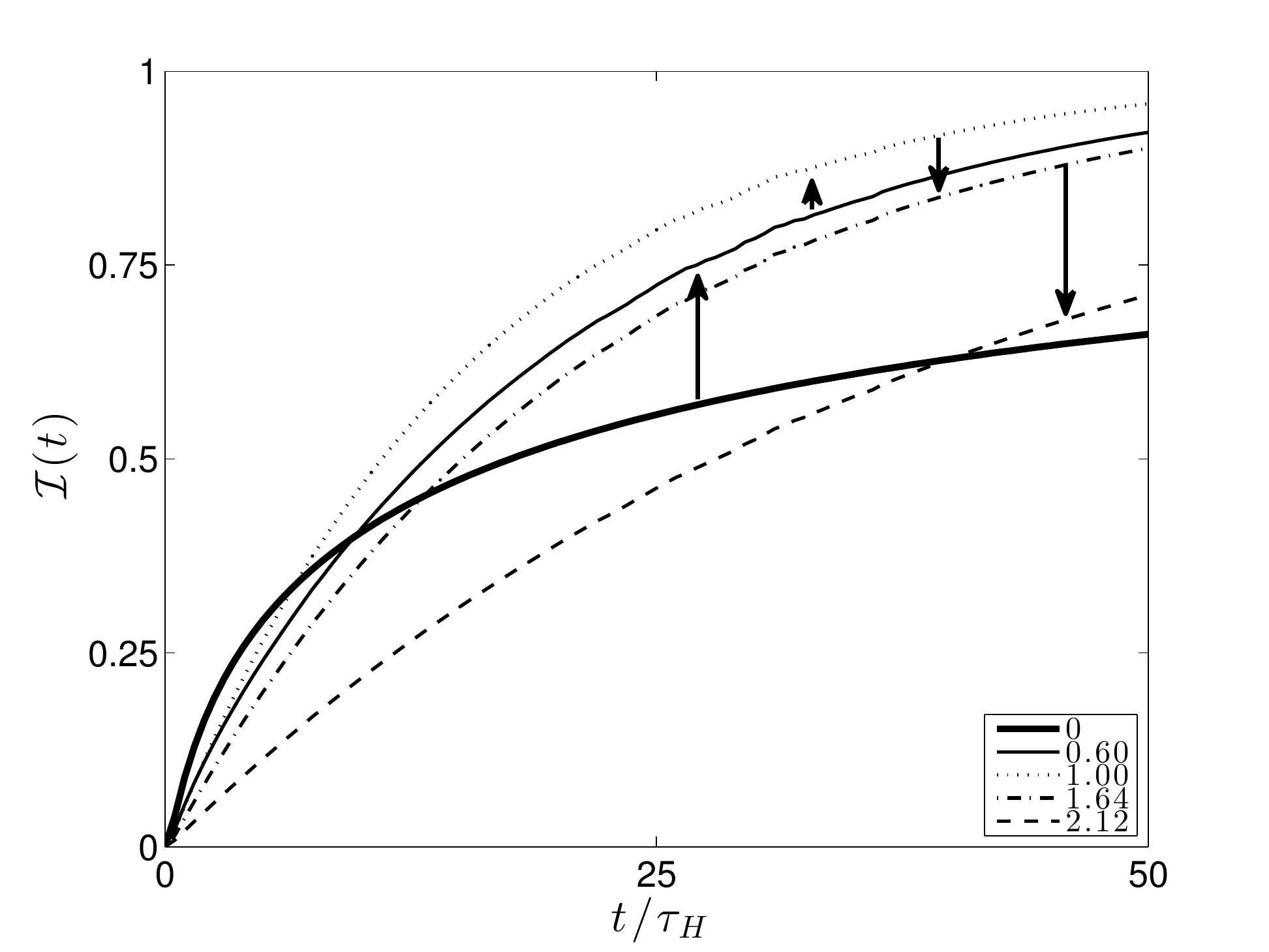}
   \caption{Non-monotonic variation of the infection curve $\mathcal{I}(t)$ with change in $\gamma$, the steepness of the potential confining the animals to their home ranges. Time is scaled to $\tau_H$; $\mathcal{C}_1$ scaled to $2D/H$ equals $0.3$. Starting with the unconfined case, increase in $\gamma$ makes infection more effective for small values of $\gamma$ but less effective for larger values. The value shown for each line in the legend is of $H/\sigma,$ the ratio of the inter-center distance to the steady-state Smoluchowski width. }
   \label{fig:Fig1}
\end{figure}

The infection curve $\mathcal{I}(t)$ is now obtained by calculating  the Laplace transforms of Eq. (\ref{numut}), substituting them in Eq. (\ref{iepsilon}), and inverting the transform. We do this with the help of a simple numerical code implemented in Matlab and verify the results by direct numerical solution of the partial differential equation (\ref{start}). See the appendix of ref. ~\cite{SSK:2013} where a similar procedure is explained in detail. The agreement is excellent except for confining potentials that are so steep that the direct numerical procedure used for verification breaks down. Our calculated $\mathcal{I}(t)$ for initial location of the animals at their home range centers, and for an assumed contact rate parameter $\mathcal{C}_1$ equal to $0.3$ in units of $2D/H$,  is displayed in  Fig.~\ref{fig:Fig1}  as a function of $t$ scaled to $\tau_H$, for various steepness values of the confining potential. Here $\tau_H=H^2/2D$ is the time required for either animal to traverse diffusively the inter-center distance, and we attach the suffix $1$ to $\mathcal{C}$ to emphasize that this result is $1d$. Striking behavior is apparent in Fig.~\ref{fig:Fig1}.

Recall that $\sigma=\sqrt{2D/\gamma}$ is the width of steady-state distribution of the Smoluchowski walker in $1d$. We keep $D$ and the inter-center distance $H$ constant, and increment $\gamma,$ thereby changing $\sigma$. The case of no confining potential corresponds to the thick solid curve ($H/\sigma=0$). We gradually increase the confinement steepness, giving the latter parameter  the respective values $0.6$ (thin solid line), $1.0$ (dotted), $1.64$ (dot-dashed) and $2.12$ (dashed). Generally, as time proceeds, $\mathcal{I}(t)$ rises from $0$ and saturates to $1$. Infection may be said to occur faster as the confining potential becomes steeper but only for relatively small values of $\gamma$. Further increases make the infection proceed slower. Vertical arrows between curves show this march graphically. Reversal in their direction marks the interesting phenomenon. This non-monotonic behavior is noteworthy, one of the primary results of our analysis, and is also observed if the diffusion constant of the animals is varied keeping the potential steepness constant. It arises from the interplay of three quantities, the diffusion constant $D$, the steepness $\gamma$ and the inter-center distance $H$ which here is also the distance between the initial locations of the animals. For a given value of $H,$ changes in $D$ or $\gamma$ uncover the phenomenon. Varying $H$ does not: maximum transmission occurs when $H=0$, i.e., when the animals do not have to move to find each other for the infection to be propagated. The key parameter is $\gamma\tau_H=H^2\gamma/2D$ which is nothing other than $\left(H/\sigma\right)^2$: for a given $H,$ optimum transmission of infection occurs when the parameter equals 1, particularly in the capture-limited case. More generally the critical value is different from 1.  
 
\emph{Reduction to the case of no confinement}--Given that many of the previous quantitative theories do not explicitly incorporate home range confinement, it is important to ask what our model calculation predicts for such free diffusion. In that case, a full analytic solution is possible. With $\gamma\rightarrow 0$,  $\nu(t)$ and $\mu(t)$ in $1d$ are simple propagators of the diffusion equation,
\begin{equation}
\nu(t) = \frac{1}{\sqrt{ 8\pi Dt}}; \quad \mu(t) = \frac{1}{\sqrt{ 8\pi Dt}}e^{-\frac{H^2}{8Dt}}.
\end{equation}
Their Laplace transforms are known. With the introduction of a time $\theta = 8D/(\pi \mathcal{C}_1^2)$ that incorporates the diffusion constant and the capture parameter, we have for the infection probability in the Laplace domain, 
\begin{equation}
  \tilde{I}(\epsilon) =\frac{1}{\epsilon} \left(\frac{e^{-\sqrt{\epsilon\tau_H}}}{1+\sqrt{\epsilon \theta}}\right).
\end{equation}
Inverse transformation  gives the analytic time domain result
\begin{equation}
I(t) = \text{erfc}\left(\sqrt{\frac{\tau_H}{4t}}\right)-e^{\left(\sqrt{\frac{\tau_H}{4t}}+\frac{t}{\theta}\right)}\text{erfc}\left(\sqrt{\frac{\tau_H}{4t}}+\sqrt{\frac{t}{\theta}}\right).
\label{oldresult}
\end{equation}
We have not encountered this result in the epidemic literature earlier. However, curiously, the expression has been reported independently by several authors in varied reaction diffusion contexts \cite{KSinJPC,ABRAMSON:1995:ID169,REDNER:2001:ID188,CARSLAW}. The further simplification of an infinite contact rate (motion limit), leading to a vanishing $\theta$, yields the simple diffusion result that the infection curve is given by a complementary error function of argument $\sqrt{\tau_H/4t}$. The time dependence of Eq.~(\ref{oldresult}) is depicted as the thick solid line $\gamma=0$ in Fig. 1. 

\emph{Effective Rates of Infection and Extension to Dense Systems}--The foregoing analysis, while exact for dilute systems, is not applicable for dense systems because they contain numerous (rather than one) interacting pairs whose dynamics, and even identity, evolve in time. We have developed, and plan to report in a forthcoming publication, an approximate kinetic equation theory applicable to such situations, along the lines of ref.~\cite{KENKRE203:233}. For use in that theory, we extract from the above single-pair analysis an effective infection rate in the same spirit as in the calculation of a Fermi Golden Rule rate for describing transitions in a complex quantum system. Inspection of Fig.~1 shows that the overall shape of the infection curve $\mathcal{I}(t)$ is \emph{similar} to an exponentially rising function $1-e^{-\alpha t}$ which could be said to correspond to an infection rate $\alpha$. In this simple case, the statement $\tilde{\mathcal{I}}(\epsilon)=\alpha/\epsilon(\epsilon+\alpha)$ would apply. Comparison with Eq.~(\ref{iepsilon}) shows that the actual infection curve corresponds to an infection \emph{memory} given in the Laplace domain by
\begin{equation}
\tilde{\alpha}(\epsilon) = \frac{\epsilon\tilde{\mu}(\epsilon)}{(1/\mathcal{C})+\tilde{\nu}(\epsilon)-\tilde{\mu}(\epsilon)}
\label{alphamemory}
\end{equation}
from which we  extract an effective rate $\alpha$ in the Markoffian limit, $\epsilon \rightarrow 0$. With the introduction of a \emph{motion parameter} $\mathcal{M}$ as the reciprocal of $\int_{0}^{\infty}dt\left[\nu(t)-\mu(t)\right]$, we get
\begin{equation}
\alpha\equiv\lim_{\epsilon\rightarrow 0}\tilde{\alpha}(\epsilon) = \frac{\mu(\infty)}{(1/\mathcal{C})+(1/\mathcal{M})}.
\label{alpharate}
\end{equation}
An Abelian theorem has been used in the last equality to express $\alpha$ in terms of quantities in the time domain.
The effective rate now appears as the product of the probability in the steady state that the two walkers occupy the same position, independently of the initial condition (essentially the numerator), and a combined rate involving the contact parameter and a motion parameter (essentially the reciprocal of the denominator). Thus, $\alpha$ equals simply $\mathcal{C}\mu(\infty)$ in the contact-limited case, i.e., when $\mathcal{C}<<\mathcal{M}$. In the opposite limit $\mathcal{M}<<\mathcal{C}$, infection is governed by the motion and $\alpha$ is $\mathcal{M}\mu(\infty)$. This is clear in the left panel of Fig.~2. The motion parameter $\mathcal{M}$ describes an accumulated integral of the difference between the two probability densities explained above of the two walkers coinciding in location. The non-monotonicity effect is displayed in the right panel of Fig. 2 where $\alpha$ rises, peaks, and drops as the potential steepness is varied. 

Equation~(\ref{crux}) allows the evaluation of $\mu(\infty)$ in Eq.~(\ref{alpharate}) for arbitrary dimensions $s$ as being $\left[(1/\sigma\sqrt{2\pi})e^{-H^2/2\sigma^2}\right]^s$ where $\sigma=\sqrt{2D/\gamma}$ is the width of steady-state distribution in $1d$.  Calculating $\mathcal{M}$ involves the evaluation of an improper integral which is convergent  in 1-d \cite{SATOMITHESIS} but presents the standard difficulties that arise in reaction diffusion problems in dimensions higher than $1$ if reaction is taken to occur at points as we have done here. Generally, generalizing the treatment to include reaction in finite regions solves this problem. It is also of interest to include the consequences of the introduction of a decay into the system. Such a decay may arise from radiative lifetimes as explained for excitons in molecular crystals earlier~\cite{KENKRE:1983}, from finite lifetimes $\tau$ of the infected animals as they may die from natural death or from predator attack, or from finite lifetime of the infection itself. The latter may be caused by the animals recovering from being infective. In such cases one takes the limit $\epsilon \rightarrow 1/\tau$ rather than $\epsilon \rightarrow 0,$ and Eq. (\ref{alpharate}) is replaced by
\begin{equation}
\alpha \tau= \frac{\int_{0}^{\infty}dt e^{-t/\tau}\mu(t)}{(1/\mathcal{C})+\int_{0}^{\infty}dt e^{-t/\tau}\left[\nu(t)-\mu(t)\right]}.\label{alphatau}
\end{equation}
In case a natural finite lifetime is absent in the given problem, it may be natural to introduce it as a probe time associated with measurement. 
\begin{figure}[h] 
  \centering
   \includegraphics[width=1\columnwidth]{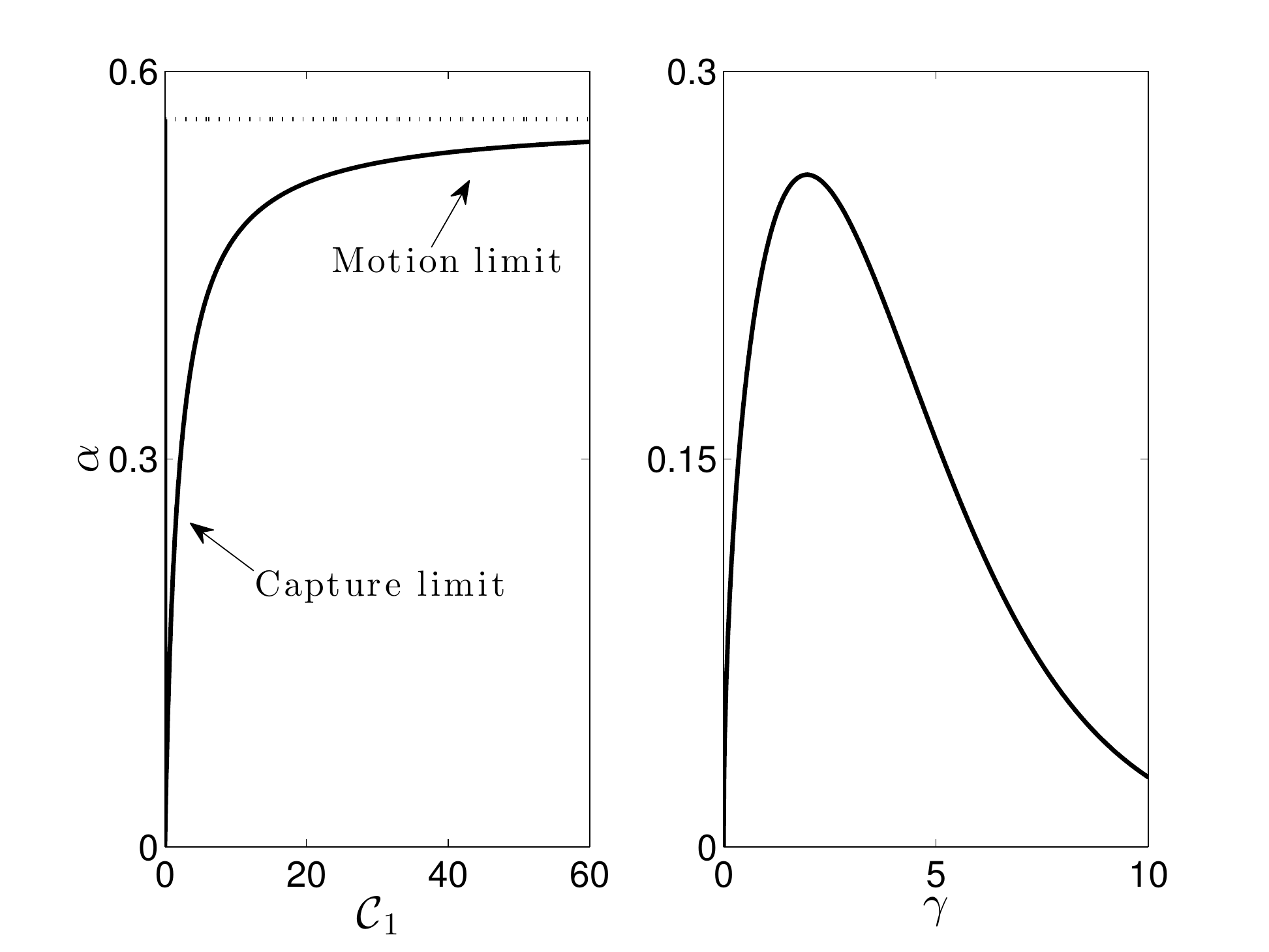}
   \caption{Dependence of the effective infection rate $\alpha$ from Eq.~(\ref{alpharate}) on the contact rate $\mathcal{C}_1$ (scaled to $2D/H$) in the left panel and on the confining potential steepness $\gamma$ in the right panel. Both $\alpha$ and $\gamma$ are scaled to $1/\tau_H$. The left panel shows that $\alpha$ is linear in the contact rate for small values of the latter but saturates to the motion-limited value ($0.56$ in this example) for large values. The right panel shows the non-monotonicity effect on infection: as confinement steepness $\gamma$ increases, $\alpha$ rises to a peak and decreases for larger $\gamma$. For the right panel, $\mathcal{C}_1$ in units of $2D/H$ is 15.}
   \label{fig:Fig2}
\end{figure}

\emph{Conclusions}--The calculation we have presented is precise for the limited model considered and is valid for movement both with and without spatial constraints imposed on the moving animals, the latter to represent the existence of home ranges. In the presence of spatial constraints, the analysis has uncovered a remarkable phenomenon: infection efficiency is non-monotonic when the steepness of the confining potential, or the animal diffusion constant, is varied~\cite{reactdiff}. Each of the two quantities thus has a critical value on both sides of which infection becomes inefficient. An understanding of the curious effect we observe can be achieved at various levels. The effect involves three quantities, the distance $H$ between the centers of the home ranges, the diffusion constant $D$, and the potential steepness $\gamma$. Combined into a single parameter $\sqrt{H^2\gamma/2D}$, which equals $H/\sigma$, the quantities signal inefficient transmission of infection when variations in $D$ or $\gamma$ make the parameter differ from its optimum value. In the capture-limited case, the optimum value is  $1$ and corresponds to the static statement that the width of the steady state distribution of the Smoluchowski equation equals the distance between the home centers; or to the dynamic statement that the time taken by the walker to traverse the inter-homerange distance $H$ diffusively equals the time $1/\gamma$ characteristic of free motion of the walker to the center under the action of the potential. Away from the capture limit, the optimum value changes from $1$ because of contributions from what has been explained as the motion parameter $\mathcal{M}$ (see earlier text). Thus, in the right panel of Fig.~\ref{fig:Fig2}, it happens to equal 1.97. The analysis is applicable for arbitrary initial conditions. In addition to being exact for the simplified model considered, it provides a sound basis for obtaining expressions for infection rates that can be used in approximate, but  practical, theories of the spread of infection. Such extended theories are appropriate in  realistic scenarios involving dense animal populations, will be reported elsewhere, and consist of a kinetic equations setup as in refs.~\cite{AK:2002} and~\cite{KENKRE203:233} but whose infection (aggression) rates are   computed from the present analysis rather than being simply postulated.The formalism is directly useful for the study of the spread of zoonotic diseases such as the Hantavirus \cite{Hanta} in which infection spreads as the result of the movement of rodents on a terrain. It should also find use in other contexts as in the study of West Nile Virus \cite{WestNile1,WestNile3} within the field of epidemics and also in general studies of reaction diffusion and interacting random walks. 

It is a pleasure for us to acknowledge  helpful conversations with Professor Kathrin Spendier of the University of Colorado. This research was supported by the Consortium of the Americas for Interdisciplinary Science and by the Program in Interdisciplinary Biological and Biomedical Sciences of the University of New Mexico.

\end{document}